# Graphdiyne-metal contacts and graphdiyne transistors

Yuanyuan Pan[1,†], Yangyang Wang[1,†], Lu Wang[4], Hongxia Zhong[1], Ruge Quhe[1,3], Zeyuan Ni[1], Meng Ye[1], Wai-Ning Mei[4], Junjie Shi[1], Wanlin Guo[5], Jinbo Yang[1,2] ∗, Jing Lu[1,2] ∗

[1]State Key Laboratory of Mesoscopic Physics and Department of Physics,

Peking University, Beijing 100871, P. R. China

[2]Collaborative Innovation Center of Quantum Matter, Beijing 100871, P. R. China

[3]Academy for Advanced Interdisciplinary Studies, Peking University, Beijing 100871,

P. R. China

[4]Department of Physics, University of Nebraska at Omaha, Omaha, Nebraska 68182-0266

[5]State Key Laboratory of Mechanics and Control of Mechanical Structures, Key Laboratory for

Intelligent Nano Materials and Devices of the Ministry of Education, and Institute of Nanoscience,

Nanjing University of Aeronautics and Astronautics, Nanjing 210016, P. R. China

[†]These authors contributed equally to this work.

∗ Corresponding authors: jinglu@pku.edu.cn, jbyang@pku.edu.cn

## ABSTRACT

Graphdiyne is prepared on metal surface, and making devices out of it also inevitably involves contact with metals. Using density functional theory with dispersion correction, we systematically studied for the first time the interfacial properties of graphdiyne contacting with a series of metals (Al, Ag, Cu, Au, Ir, Pt, Ni, and Pd). Graphdiyne is in an *n*-type Ohmic or quasi-Ohmic contact with Al, Ag, and Cu, while it is in a Schottky contact with Au (at source/drain interface), Pd, Pt, Ni, and Ir (at source/drain-channel interface), with high Schottky barrier heights of 0.39, 0.21 (*n*-type), 0.30, 0.41, and 0.45 (*p*-type) eV, respectively. A graphdiyne field effect transistor (FET) with Al electrodes is simulated by using quantum transport calculations. This device exhibits an on-off ratio up to $10^4$ and a very large on-state current of $1.3 \times 10^4$ mA/mm in a 10 nm channel length. Thus, a new prospect is opened up for graphdiyne in high performance nanoscale devices.

Due to three hybridization states ($sp^1$, $sp^2$ and $sp^3$), carbon can form a set of allotropes, such as fullerenes,[1] carbon nanotubes,[2] and graphene.[3] Graphene has significant potential application in the nanoelectronics because of its high carrier mobility. But the zero band gap limits its application in effective field effect transistor (FET). In spite of the fact that additional efforts such as applying an electric field or single-side adsorption of atoms/molecules on bilayer and ABC-stacked few-layer graphene[4] or sandwiching graphene by BN single layer have been done,[5] opening a large band gap without degrading its electronic properties remains a tough challenge for graphene. As a novel two-dimensional carbon allotrope that has both $sp^1$ and $sp^2$ carbon atoms,[6,7] graphdiyne was first prepared by Li's group on the Cu surface via a cross-coupling reaction using hexaethynylbenzene in 2009.[8] It has a conductivity of $2.516 \times 10^{-4}$ S/m, typical of a semiconductor. The calculated band gap at the density functional theory (DFT) level is about 0.5 eV.[9] The calculated in-plane intrinsic electron and hole mobility in graphdiyne can reach the order of $10^5$ and $10^4$ $cm^2V^{-1}s^{-1}$,[10] respectively, at room temperature, which are comparable with those of graphene. As one atomic layer thick material, short-channel effects are expected to be greatly suppressed in graphdiyne,[11] and a FET based on graphdiyne probably can be scaled down to very short channel length. Therefore graphdiyne is a promising candidate material for high-speed applications in logic devices. To date, most previous studies about graphdiyne have focused mainly on the electronic, optical, and mechanical properties,[9,12,13,14] or the application in hydrogen purification and storage,[15] solar cells,[16] photocatalytics,[17] anode of lithium batteries,[18,19] etc. Very little attention was paid to its FET performance.

As we know, fabricating devices out of graphdiyne inevitably involves contact with metal electrodes, and good contact always improves device performance remarkably. Together with the fact that graphdiyne is prepared on the Cu substrate, it is of fundamental interest to explore the graphdiyne-metal contacts from theoretical aspect. However, to the best of our knowledge, the interfacial properties of graphdiyne on metals remain an open question.

In this paper, we systematically study for the first time the interfacial properties of graphdiyne on a variety of metals (Al, Ag, Cu, Au, Ir, Pt, Ni, and Pd) by using density functional theory with dispersion correction. The contact of graphdiyne with Al, Ag, Cu, Au,

Ir, and Pt is a weak physisorption, while with Ni and Pd is a strong chemisorption. The electronic structure of graphdiyne is strongly perturbed by Ir, Pt, Ni, and Pd contacts. Graphdiyne is *n*-type doped when contacted with Al, Ag, Cu, Au, and Pd electrodes, but *p*-type doped when contacted with Ir, Pt, and Ni electrodes. The contact of graphdiyne with Al, Ag, Cu electrodes is in an Ohmic or quasi-Ohmic contact, while it with Au, Pd, Pt, Ni, and Ir electrodes is in a Schottky contact, and Schottky barriers is 0.39, 0.21, 0.30, 0.41, and 0.45 eV, respectively. According to Schottky barriers and tunneling barrier, five types of metal-graphdiyne contacts are formed. Subsequently, a graphdiyne FET device with Al electrodes is designed. The electron transport properties are calculated by using the quantum transport approach. The 10 nm-channel-length graphdiyne FET reveals an on-off ratio up to $10^4$, suggestive of the great potential of graphdiyne as the channel of a high performance nanoscale FET.

**Results and Discussion**

**Graphdiyne-metal contacts**

Three kinds of initial configurations are chosen for graphdiyne on Au surfaces (the center of the carbon hexagon of graphdiyne on the top of metal atoms of A, B, and C layer, respectively). After relaxation, we find that the most stable configuration of graphdiyne on Au surfaces is the one with the center of the carbon hexagon of graphdiyne on the top of A-layer metal atoms, as shown in Fig. 1. The other two configurations are 0.15 and 0.14 eV/supercell higher in energy, respectively. Graphdiyne on other metals surfaces adopts the same configuration as that on Au surfaces.

The key interfacial structure and properties parameters of graphdiyne-metal contacts are listed in Table 1. The binding energy $E_b$ of the graphdiyne-metal contact is defined as

$$E_b = (E_G + E_M - E_{G/M}) / N \tag{3}$$

where $E_G$, $E_M$, and $E_{G/M}$ are the relaxed energy for graphdiyne, the clean metal surface, and the combined system, respectively, and $N$ is the number of interfacial carbon atoms in a supercell. The interfacial distance $d_{C-M}$ is defined as the average distance from the innermost layer of metal to graphdiyne surfaces.

The metal-graphdiyne interfacial structures can be classified into two classes in terms of the binding energy $E_b$ and equilibrium distance $d_{C-M}$. Adsorption of graphdiyne on Ni and Pd(111) surfaces is chemisorption with $E_b > 0.23$ eV and $d_{C-M} < 2.25$ Å. In contrast, adsorption of graphdiyne on Al, Ag, Cu, Au, Ir, and Pt(111) surfaces is physisorption with $E_b < 0.12$ eV and $d_{C-M} > 2.8$ Å. The classificatory standard of graphdiyne adsorbed on metals is similar to that of graphene adsorbed on metals.[20,21] There are distinct buckling heights of graphdiyne on the interfaces, as shown in Fig. 5. The buckling heights of graphdiyne adsorbed on Au, Ag, Al, Cu, and Ir are smaller with values of 0.01, 0.02, 0.05, 0.09, and 0.13 Å, respectively, while on Ni, Pd, and Pt are larger with values of 0.58, 0.70, and 0.88Å, respectively. As for the metal surface, the buckling of the A layer of Ir, Cu, Al, and Ag is small with buckling heights less than 0.13 Å, while there is apparent buckling of the A layer for Pt, Ni, Pd, and Au with buckling heights of 0.36, 0.37, 0.43, and 0.73 Å, separately.

The electronic structures of graphdiyne-metal contacts are plotted in Fig. 2. In terms of the hybridization degree of graphdiyne on metals, the band structure of metal-graphdiyne contact is classified into two categories. The band structure of graphdiyne is identifiable clearly for graphdiyne on Al, Ag, Cu, and Au surfaces, as a result of weak charge-transfer interaction and dispersion interaction between graphdiyne and Al, Ag, Cu, and Au surfaces. The band structure of graphdiyne absorbed on Al is less intact than that on Ag, Cu, and Au, because the outmost 3$p$ orbitals of Al is partially occupied, which slightly hybridizes with the states of graphdiyne above the Fermi level ($E_F$), whereas the outmost electrons of the other three metals are all partially filled $s$ states. The band gap of graphdiyne on Al, Cu, Au, and Ag surfaces are 0.31, 0.36, 0.39, and 0.47 eV, respectively, compared with a band gap of 0.46 eV in free-standing graphdiyne. The band structure of graphdiyne is destroyed seriously for graphdiyne on Ir, Pt, Pd, and Ni surfaces, because the outmost $d$ electrons of the four metals are strongly hybridize with the states near $E_F$ of graphdiyne.

The total electron distributions in real space of Ag-graphdiyne and Pd-graphdiyne interfaces are compared in Fig. 3. There is no electron accumulation between Ag and graphdiyne surface, suggesting absence of covalent bond between them. By contrast, electrons are accumulated between Pd and graphdiyne surfaces, indicating the formation of covalent bond between them. The difference confirms that the adsorption of graphdiyne on Ag is physisorption, with the band structure of graphdiyne intact, whereas that on Pd is chemisorption, with the band structure of graphdiyne distorted seriously.

Schottky barrier, tunneling barrier, and Fermi level pinning play important roles in a FET. The schematic diagram of a graphdiyne FET is shown in Fig. 4a. Schottky barrier can appear on two different interfaces in a graphdiyne FET: One is between graphdiyne and the contacted metal surface in the vertical direction (labeled interface B, and the corresponding Schottky barrier is labeled $\Phi_V$), and the other is between the contacted and the channel graphdiyne in the lateral direction (labeled interface D, and the corresponding Schottky barrier is labeled $\Phi_L$).[22] Other type of barrier can also occur at interface D if the band position between the contacted and the channel graphdiyne is different. Tunneling barrier can appear at interface B when electrons cross the gap (normally van der Waals gap) between metal and graphdiyne.

The absolute band position is an important metric to evaluate the Schottky barriers in graphdiyne FET. It is well known that the band gap of a semiconductor is seriously underestimated in DFT method (The band gap of graphdiyne calculated by DFT and *GW* methods is 0.46 and 1.1 eV,[9] respectively.), and a *GW* correction can give a band gap consistent with photoemission/inverse photoemission gap measurements. Therefore, a *GW* correction is necessary to obtain the correct band position. We propose that the Fermi level or the band gap center (BGC), namely, the average at the valence band maximum (VBM) and the conduction band minimum (CBM) is unchanged after *GW* correction, and the energy at CBM and VBM of graphdiyne after *GW* correction $E_C^{GW}$ and $E_V^{GW}$ can then be obtained as

$$E_C^{GW} = E_F + \frac{1}{2} E_g^{GW} \qquad (4)$$

$$E_C^{GW} = E_F - \frac{1}{2} E_g^{GW} \qquad (5)$$

Where $E_g^{GW}$ is the band gap of graphdiyne by *GW* approach. $E_F$ is the Fermi level of graphdiyne obtained by DFT method. Fig. 4b illustrates *GW* correction to the absolute band position. Such a correction scheme based on the unchanged Fermi level (termed *GW*–BGC scheme) has also been adopted to obtain the absolute band position by Jiang[23], Gong *etc.*[24]

and Toroker *etc*.[25]. The calculated ionization potential (IP = $-E_V^{GW}$ = 5.45 eV) and electron affinity ($\chi = -E_C^{GW}$ = 4.22 eV) of bulk MoS$_2$ by *GW*–BGC scheme,[23] compared with values of IP = 5.33 and $\chi$ = 4.45 eV from DFT– perdew-Burke-Ernzerhof (PBE) functional,[23] are in good agreement with the experimental values (IP = 5.47 ±0.15 eV and $\chi$ = 4.07 ±0.35 eV).[26]

Vertical Schottky barrier $\Phi_V$ is determined by the band structures of graphdiyne underneath metals and the Fermi level $E_F$ of absorbed systems shown in Fig. 2. Because $E_F$ of the absorbed systems is above the CBM of graphdiyne on Al, Ag, and Cu surfaces at both DFT and *GW*-BGC schemes, there is no vertical metal-graphdiyne Schottky barrier for the three contacts at the two schemes. The Ohmic contact between graphdiyne and Al and Cu surfaces in the vertical direction has been measured by Li's group.[8] Nor does a vertical Schottky barrier exist for Ir, Pt, Ni, and Pd contacts at the two schemes because a strong band hybridization has taken place. By contrast, there is a quite small vertical Schottky barrier in terms of the difference between the $E_F$ of absorbed systems and the CBM of graphdiyne underneath Au electrodes of $\Phi_V^{DFT}$ = 0.01 eV at DFT scheme while it is corrected to a large value of $\Phi_V^{GW}$ = 0.39 eV at *GW*–BGC scheme.

Lateral Schottky barrier $\Phi_L$ is determined by the energy difference between the absorbed system Fermi level and the CBM (*n*-type) or the VBM (*p*-type) of channel graphdiyne. Graphdiyne forms an Ohmic contact with Al, Ag, Cu, and Au in the lateral direction at DFT method since the CBM of graphdiyne underneath metal is higher than that of channel graphdiyne at DFT method; After *GW*-BGC correction, the lateral Ohmic contact remains for Al and Au contact but a rather small Schottky barrier appears for Ag and Cu contacts with $\Phi_L^{GW}$ = 0.02 and 0.08 eV, respectively. Such a small Schottky barrier is probably blurred by thermionic emission at room temperature, and graphdiyne thus forms a quasi-Ohmic contact with Ag and Cu in the lateral direction. The Fermi level of graphdiyne underneath metal is higher than CBM or lower than VBM of channel graphdiyne at DFT scheme, so Pd and Pt form lateral Ohmic contact with graphdiyne. After *GW*-BGC correction, a high lateral Schottky barrier is formed for Pd and Pt contacts, with $\Phi_L^{GW}$ = 0.21 (*n*-type) and 0.30 (*p*-type) eV, respectively. There is a lateral p-type Schottky barrier for Ni and Ir-graphdiyne contacts at both DFT and *GW*-BGC schemes, with $\Phi_L^{DFT}$ = 0.09 (Ni) and 0.13 (Ir) eV and $\Phi_L^{GW}$ = 0.41 (Ni) and 0.45 (Ir) eV. In our calculations, the Schottky barrier directly calculated by DFT method is always much smaller than that by *GW*-BGC scheme because the band gap of graphdiyne calculated by DFT scheme of 0.46 eV are much smaller than that by *GW*-BGC scheme of 1.10 eV and the BGC of graphdiyne is assumed unchanged in the two schemes.

Fig. 5 shows the potential profiles at the vertical metal-graphdiyne interfaces. There is an obvious tunneling barrier (3.54 - 4.54 eV) at the metal-graphdiyne physisorption interfaces while there is a small (0.83 eV) or vanishing one at the metal-graphdiyne chemisorption interfaces. Similar results are found at the vertical metal-graphene interfaces.[27] Furthermore, we assume a square potential barrier to replace the real potential barrier, and the barrier height ($\Delta V$) and width ($w_B$) of the square potential barrier are the barrier height and full width at half maximum (FWHM) of the real potential barrier shown in Fig. 5. The tunneling probabilities $T_B$ is calculated using equation:[27]

$$T = \exp\left(-2 \times \frac{\sqrt{2m\Delta V}}{\hbar} \times w_B\right) \qquad (6)$$

where $m$ is the massive of free electron and $\hbar$ is reduced the Plank's constant. The resulting tunneling possibilities at Au, Cu, Ag, Al, Ir, Pt, Ni, and Pd-graphdiyne interface are 4.79%, 6.36%, 6.83%, 7.15%, 8.11%, 21.55%, 71.35%, and 100%, respectively. The tunneling possibilities of chemisorption are much larger than those of physisorption.

In terms of the Schottky barrier and tunneling barrier, five types of metal-graphdiyne contacts are identified and shown in Fig. 4c. In Type 1 contact, graphdiyne is in an Ohmic contact with Al electrodes with a tunneling barrier at interface B, and in Type 2 contact, graphdiyne is in a quasi-Ohmic contact with Ag and Cu electrodes of a quite small $\Phi_L$ at interface D and a tunneling barrier at interface B. Since work function of Au is larger than those of Al, Ag, and Cu, electrons confront a high Schottky barrier $\Phi_V$ at interface B in addition to a tunneling barrier, leading to Type 3 contact. Pd, Ir, Ni, and Pt can form covalent bond with graphdiyne and lead to the metallization of graphdiyne layer under them, thus eliminating the Schottky barrier $\Phi_V$ at interface B. In Pd-graphdiyne contact, there is only a Schottky barrier $\Phi_L$, resulting in a low-resistance Type 4 contact. In Ir, Ni, and Pt-graphdiyne contacts, electrons traverse a tunneling barrier at interface B and then confront a high Schottky barrier at interface D, forming Type 5 contact.

Fig. 6 shows the line-up of metal Fermi level with the electronic bands of graphdiyne after *GW* correction. There is no obvious Fermi level pinning in metal-graphdiyne contacts, while partial Fermi level pinning is calculated in MoS$_2$-metal contact[28,29] and graphdiyne-metal contact.[30] The Fermi levels of Al, Ag, Cu, Au, and Pd-graphdiyne absorbed systems are higher than that of the channel graphdiyne and form *n*-type contact, while the Fermi levels of Ir, Pt, and Ni-graphdiyne absorbed systems are lower than that of their channel graphdiyne and form *p*-type contact. Therefore, graphdiyne *p - n* junction can be fabricated by using Al, Ag, Cu, Au, or Pd to contact one end of graphdiyne and Ir, Pt, and Ni to contact the other end of it. Photoelectronic applications of graphdiyne can be developed.

**Graphdiyne field effect transistor**

To assess the electron transport performance of graphdiyne, we further simulate a graphdiyne FET in a top-gated two-probe model. We adopt Al as the electrodes in the following transport simulations because Al provides an Ohmic contact with graphdiyne in the vertical and lateral interface direction and Al electrode has actually be used.[8] The schematic model is presented in Fig. 4a and the distance between the Al lead and graphdiyne is 3.41 Å according to our DFT results. The dielectric region is made of SiO$_2$ with a thickness of 10 Å. To start with, we calculated the band structures of graphdiyne and the transmission spectrum of a 6 nm-channel-length graphdiyne FET using the DFT method with single-ζ (SZ) basis set to benchmark our semiempirical (SE) extended Hückel results (Fig. S1). The band structure and transmission spectrum calculated between the two methods are similar, especially the size and position of the band gap and transmission gap are highly consistent. Thus the SE approach is reliable and could be a good substitution of DFT in our transportation simulation. Then we focus on the transport properties of the graphdiyne FET with a larger channel length $L$ = 10 nm. The transmission spectra of the device under $V_g$ = 0 V is shown in Fig. 7a. A transport gap of 0.47 eV appears below $E_F$ and electrons are the majority charge carriers in this transistor, as expected from the calculated band alignment shown in Fig. 6. Therefore by applying a negative gate voltage to the channel, the conductance can be decreased, and an on-off switch is expected. The transmission coefficient of a FET $T(E)$, is proportional to the product of the projected density of states (PDOS) of electrodes and channel:[31]

$$T(E) \propto g_{ch}(E) g_L(E) g_R(E) \qquad (7)$$

where $g_{ch}(E)$ and $g_{L/R}(E)$ are the PDOS of the channel and the left/right lead, respectively. Therefore the transmission gap originates from the similar-sized PODS gap of the graphdiyne channel (Fig. 7b) and further originates from the band gap of infinite graphdiyne (0.46 eV).

As shown in Fig. 7a, at $V_g = 0$ and −5 V an obvious transport gap of about 0.47 eV is located below $E_F$, leading to a large transmission coefficient at $E_F$ ($T(E_F)$), while at $V_g = -7.1$ V, the gap is shifted to $E_F$, leading to a drastic decrease of $T(E_F)$. Fig. 7c shows the zero-bias transfer characteristics of the 10 nm-channel-length graphdiyne FET ($\sigma$ vs. $V_g$), with an apparent switching effect. The conductance decreases with increasing negative gate potential within $V_g = 0 \sim -0.71$ eV, typical of an $n$-type FET. The curve minimum, or the off state, is located at $V_g = -7.1$ V. If $V_g = -5$ V is chosen as the on-state, the on-off ratio can reach $10^4$, which already satisfies the demand of FET used in complementary metal-oxide-semiconductor-like logic and is two orders of magnitude larger than the maximum on-off ratios obtained in the recently reported dual-gated bilayer[32] and ABC-stacked trilayer[33] graphene FET experiments. The steepest sub-threshold swing (SS) is 117 mV/dec. Although the SS is higher, it can be reduced by fabricating FET with a thin high-κ dielectric film ($Al_2O_3$ or $HfO_2$). The difference in the transport properties between the on- and off-state is also reflected from a difference of the transmission eigenchannel at $E = E_F$ and at the (0, 0) point of the $k$-space, as shown in Fig. 7d. The transmission eigenvalue at $E = E_F$ and (0, 0) $k$-point under is 0.488 under $V_g = -5$ V, and the incoming wave function is scattered little and most of the incoming wave is able to reach to the other lead. By contrast, the transmission eigenvalue at this point nearly vanishes ($1.836 \times 10^{-5}$) under $V_g = -7.1$ V, and the incoming wave function is nearly completely scattered and unable to reach to the other lead.

Fig. 7e presents the output characteristics of the 10 nm-channel-length graphdiyne FET acquired under $V_g = 0$ (the drain current density versus the bias voltage $J_{ds} - V_{bias}$). The linear behavior within 0.4 V bias indicates the formation of Ohmic contact between graphdiyne and Al leads, which is consistent with our DFT band calculation and previous experiments.[8] As this device is in two-contact configuration, we can estimate the upper limit for the contact resistance by $R_c = \dfrac{1}{2T_B} \dfrac{dV_{bias}}{dJ_{ds}}\bigg|_{V_{bias}=0} = 223$ Ω·μm. Although the small $T_B$ (~7.15%) in Al-graphdiyne contact which may induce a relatively larger contact resistance, the current density of the 10-nm-channel-length graphdiyne FET at $V_{bias} = 0.4$ V and $V_g = 0$ is as high as $1.3 \times 10^4$ mA/mm and already satisfies the requirement of 1480 mA/mm for the high performance FETs of 2020 from the 2013 edition of the International Technology Roadmap for Semiconductors (ITRS).[34] The large on-current in graphdiyne FETs is attributed to the high carrier mobility of graphdiyne[10] ($J \propto \mu$) and is beneficial to shorten the gate delay and speed up the device operation. When $V_{bias}$ is further increased, this 10 nm-channel-length graphdiyne FET exhibits a negative differential resistance (NDR) behavior, with a peak-to-valley ratio of 3 although no NDR is observed for long channel graphdiyne film.[8]

Computing technology requires a FET with channel length smaller than 10 nm in next decades, but bulk Si FET for some time will not perform reliably at sub-10 nm channel length, because short-channel effects are becoming more and more apparent, resulting in serious degradation of device performance and invalidation of Moore's law.[35] Graphdiyne is a possible substitute material for bulk Si at 10 nm channel length.[36] However, when the channel length decreases to 6 nm, the performance of the graphdiyne FET is greatly degraded: the on-off ratio is $10^2$, and the steepest SS is 285 mV/dec (Fig. 7c). Therefore, the short-channel effect still remarkably affects the performance of sub-10 nm graphdiyne FETs. Besides, NDR behavior is also obtained in the 6 nm-channel-length graphdiyne FET, with a peak-to-valley ratio of 2. NDR can be applied in high frequency switches, oscillators, and memories, etc.

The transmission spectra of the 10 nm-channel-length graphdiyne FET as a function of $V_{bias}$ are provided in Fig. 8a to give an insight into the observed NDR behavior. With the increasing $V_{bias}$, the transmission spectrum is shifted towards a higher energy. When $V_{bias} < 0.6$

V, the change of the transmission spectrum is insignificant. When $V_{bias} > 0.6$ V, the transmission coefficients are suppressed in both the valence and conductance bands, and another transport gap $\Delta_L$ occurs below the already existed gap $\Delta_R$, with a roughly same size with $\Delta_R$. Moreover, the gap $\Delta_R$ is elevated into the bias window. Simplified band diagram of the FET is provided in Fig. 8b-8d to illustrate the change of the transmission spectra with the increasing $V_{bias}$. When $0 < V_{bias} < 0.6$ V the band profile of the channel is merely slightly affected (Fig. 8c) and so is the transmission spectrum. According to Eq. 2, the current increases with $V_{bias}$. As $V_{bias}$ further increases, the electric potential difference between the two parts of the channel near the source and drain regions gets larger (Fig. 8d). As a result, the transport gaps induced by the band gaps in the two ends of the channel separate from each other ($\Delta_L$ and $\Delta_R$), and the gap $\Delta_R$ is elevated into the bias window by the drain voltage. The tunneling across the middle part of the channel induces a much suppressed transmission hump between the two gaps $\Delta_L$ and $\Delta_R$. The cause why the transmission coefficients above the gap $\Delta_R$ (namely in the conduction band) is suppressed remains open. Because the gap $\Delta_R$ is moved into the bias window and the non-zero transmission probabilities in the bias window (above the gap $\Delta_R$) are suppressed, the current starts to decrease, causing the NDR phenomenon.

In conclusion, we present the first systematic first-principles investigation on the interfacial properties of graphdiyne on a various metal substrates. According to the adsorption strength and electronic structures, the contact of graphdiyne with Al, Ag, Cu, Au, Ir, and Pt is a weak physisorption, with the electronic structure of graphdiyne preserved for Al, Ag, Cu and Au contacts, but destroyed for Ir and Pt contacts. The contact of graphdiyne with Ni and Pd is a strong chemisorption, with strong band hybridization occurring. Graphdiyne is in an Ohmic or quasi-Ohmic contact with Al, Ag, and Cu, while a Schottky contact with Au, Pd, Pt, Ni, and Ir, with Schottky barrier heights of 0.39, 0.21, 0.30, 0.41, and 0.45 eV, respectively. An *ab initio* quantum transport simulation is performed for a gated two-probe model made of graphdiyne contacted with Al electrodes, and a high current on-off ratio of $10^4$ and a very large on-state current of $1.3 \times 10^4$ mA/mm are obtained. This fundamental study not only provides a deep insight into graphdiyne/metal contact but also reveals high performance of graphdiyne-based devices.

**Methods**

We use five layers of metal atoms (Al, Ag, Cu, Au, Ir, Pt, Ni, and Pd) in (111) orientation to simulate the metal surface and construct a supercell with graphdiyne absorbed on one side of the metal surface, as shown in Fig. 1. We fix in-plane lattice constant of graphdiyne to the value $a = 9.45$ Å.[37] The $2\sqrt{3} \times 2\sqrt{3}$ unit cells of Al, Ag, Cu, Au, Pt, and Pd (111) and the $4 \times 4$ unit cells of Ir and Ni (111) faces are adjusted to graphdiyne $1 \times 1$ unit cell, respectively. The approximation is reasonable since the metal surface constant mismatch is 0.96 ~ 8.2% with that of graphdiyne. A vacuum buffer space of at least 15 Å is set. Graphdiyne mainly interacts with the topmost two layers metal atoms, so cell shape and the bottom three layers of metal atoms are fixed.

The geometry optimizations are performed with the ultrasoft pseudopotentials plane-wave basis set with cut-off energy of 240 and 310 eV separately, implemented in the CASTEP code.[38] Generalized gradient approximation[39] (GGA) of PBE form to the exchange-correlation functional is used. To interpret the dispersion interaction among graphdiyne, a DFT-D SE dispersion-correction approach is adopted.[40] To obtain reliable optimized structures, the maximum residual force is less than 0.01 eV/Å and energies are converged to within $5 \times 10^{-6}$ eV per atom. The Monkhorst-Pack $k$-point mesh[41] is sampled with a separation of about 0.02 Å$^{-1}$ in the Brillouin zone during the relaxation and electronic calculation periods. The electronic structure calculations are analyzed via additional calculations based on the plane-wave basis set with a cut-off energy of 400 eV and the projector-augmented wave (PAW) pseudopotential[42] implemented in the Vienna *ab initio* simulation package (VASP) code.[43,44]

Transport properties are calculated by using SE extended Hückel method coupled with nonequilibrium Green's function (NEGF) method, which are implemented in Atomistix Tool Kit (ATK) 11.2 package.[45,46,47] Hoffman basis set is used, the real-space mesh cutoff is 270 eV, and the temperature is set at 300 K. The electronic structures of electrodes and central region are calculated with a Monkhorst–Pack[41] $50 \times 1 \times 50$ and $50 \times 1 \times 1$ $k$-point grid, respectively. The zero-bias conductance and the current at a finite bias are given by the Landauer-Büttiker formula:[48]

$$\sigma(V_g) = \frac{2e^2}{h} \int T_{V_g}(E, V_{bias}=0) f'(E-E_f) dE \approx \frac{2e^2}{h} T_{V_g}(E_f, V_{bias}=0) \quad (1)$$

$$I(V_g, V_{bias}) = \frac{2e}{h} \int_{-\infty}^{+\infty} \{T_{V_g}(E, V_{bias})[f_L(E-\mu_L) - f_R(E-\mu_R)]\} dE \quad (2)$$

where $T_{V_g}(E, V_{bias})$ is the transmission probability at a given gate voltage $V_g$ and bias voltage $V_{bias}$, $f_{L/R}$ the Fermi-Dirac distribution function for the left (L)/right (R) electrode, and $\mu_{L/R}$ the electrochemical potential of the L/R electrode, and $\mu_{L/R} = E_F \pm V_{bias}/2$. GGA) of PBE form[39] to the exchange-correlation functional is used through this paper.

**Acknowledgment**

This work was supported by the National Natural Science Foundation of China (No. 11274016), the National Basic Research Program of China (No. 2012CB619304), Fundamental Research Funds for the Central Universities, National Foundation for Fostering Talents of Basic Science (No. J1030310/No. J1103205), Program for New Century Excellent Talents in University of MOE of China, and Nebraska Research Initiative (No. 4132050400) and DOE DE-EE0003174 in the United States. Parts of the calculations are performed at the University of Nebraska Holland Computing Center.

**Author contributions**

The idea was conceived by J. L. The calculation was performed by Y. P. and Y. W. The data analyses were performed by Y. P., Y. W., L. W., H. Z. and J. L. Z. N., M. Y., W. M., J. S. and W. G. took part in discussion. This manuscript was written by Y. P., Y. W., J. Y., and J. L. All authors reviewed this manuscript.

**Additional information**

Supplementary information accompanies this paper at http://www.nature.com/Scientificreports.

Competing financial interests: The authors declare no competing financial interests.

**Table 1.** Calculated interfacial properties of graphdiyne on metal substrates. $a$ represents the cell parameters of the surface unit cells for various metals. $\Delta a$ is the metal surface constant mismatch. The equilibrium distance $d_{C-M}$ is the distance between the carbon atoms of graphdiyne and the relaxed positions of the topmost metal layer in the $z$ direction. The binding energy $E_b$ is the energy of per carbon atom to remove graphdiyne from metal surfaces. $W_M$ and $W$ are the calculated work function for clean metals surface and adsorbed graphdiyne respectively. $E_g$ is the band gap of graphdiyne, $\Phi_V^{DFT}$ ($\Phi_V^{GW}$) and $\Phi_L^{DFT}$ ($\Phi_L^{GW}$) are the Schottky barrier in vertical and lateral direction by DFT ($GW$-BGC) methods, respectively. $\Delta V$, $w_B$, and $T_B$ are tunneling barrier height, tunneling barrier width, and tunneling possibility, respectively. The calculated work function of graphdiyne is $W_G = 5.14$ eV, which is much larger that a value of 4.5 eV for graphene.[20,21,4]

|  | Al | Ag | Cu | Au | Ir | Pt | Ni | Pd |
|---|---|---|---|---|---|---|---|---|
| $a$ (Å) | 9.92 | 10.01 | 8.85 | 9.99 | 10.22 | 9.61 | 9.97 | 9.53 |
| $\Delta a$ (%) | 4.95 | 5.89 | 6.31 | 5.70 | 6.23 | 1.70 | 5.46 | 0.96 |
| $d_{C-M}$ (Å) | 3.41 | 3.40 | 3.22 | 3.45 | 3.11 | 2.88 | 2.24 | 2.18 |
| $E_b$ (eV) | 0.12 | 0.09 | 0.11 | 0.09 | 0.10 | 0.11 | 0.32 | 0.23 |
| $W_M$ (eV) | 4.07 | 4.66 | 4.63 | 5.10 | 5.53 | 5.20 | 5.26 | 5.24 |
| $W$ (eV) | 4.35 | 4.61 | 4.64 | 5.05 | 5.24 | 5.39 | 5.28 | 4.80 |
| $E_g$ (eV) | 0.31 | 0.47 | 0.36 | 0.40 | 0 | 0 | 0 | 0 |
| $\Phi_V^{DFT}$ (eV) | 0 | 0 | 0 | 0.01 | 0 | 0 | 0 | 0 |
| $\Phi_V^{GW}$ (eV) | 0 | 0 | 0 | 0.39 | 0 | 0 | 0 | 0 |
| $\Phi_L^{DFT}$ (eV) | 0 | 0 | 0 | 0 | 0.13 | 0 | 0.09 | 0 |
| $\Phi_L^{GW}$ (eV) | 0 | 0.02 | 0.08 | 0 | 0.45 | 0.3 | 0.41 | 0.21 |
| $\Delta V$ (eV) | 3.97 | 3.54 | 3.78 | 4.54 | 4.20 | 3.55 | 0.83 | 0 |
| $w_B$ (Å) | 1.29 | 1.39 | 1.38 | 1.39 | 1.20 | 0.80 | 0.36 | 0 |
| $T_B$ (%) | 7.15 | 6.83 | 6.36 | 4.79 | 8.11 | 21.55 | 71.35 | 100.00 |

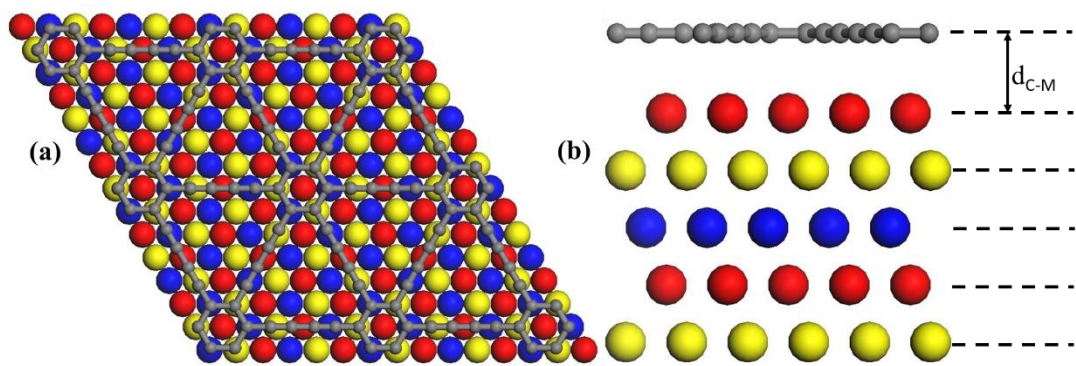

Figure 1 | (a) Top and (b) side views of the most stable configuration for graphdiyne (the gray balls) on metal surfaces (Colored balls).

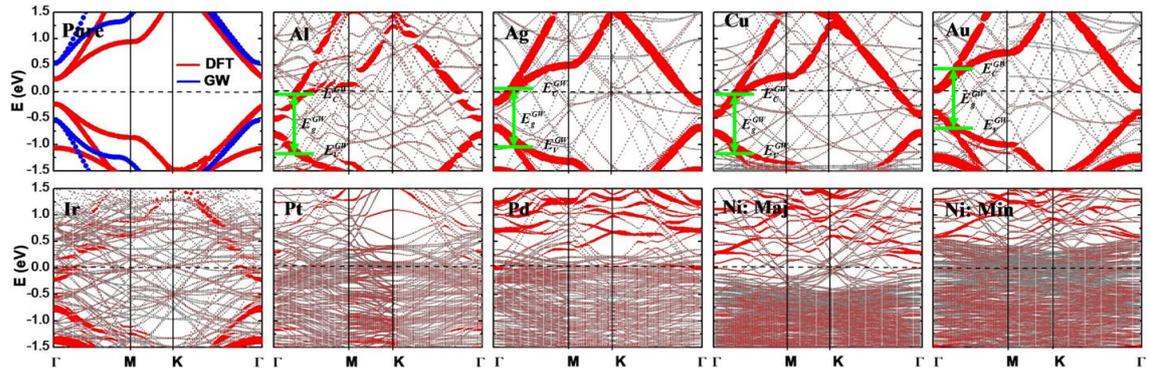

Figure 2 | Band structures of pristine graphdiyne (by DFT and *GW* methods) and graphdiyne adsorbed upon Al, Ag, Cu, Au, Ir, Pt, Pd, and Ni substrates by DFT method. The Fermi level is set at zero energy. Gray line: the bands of adsorbed systems; red line: the bands of graphdiyne; green line: the positions of CBM and VBM of graphdiyne after *GW*-BGC scheme. The line width is proportional to the weight. The labels Maj/Min indicate the majority-spin and minority-spin bands of graphdiyne on Ni substrate.

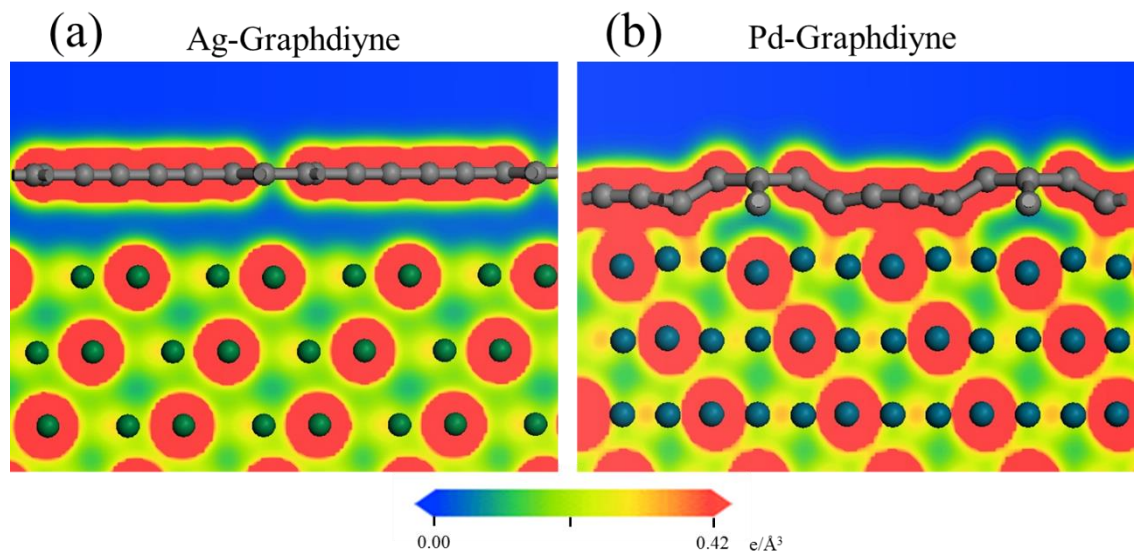

Figure 3 | Contour plots of total electron distribution of (a) Ag-graphdiyne and (b) Pd-graphdiyne interfaces. Grey, green, and blue balls are C, Ag, and Pd atoms, respectively.

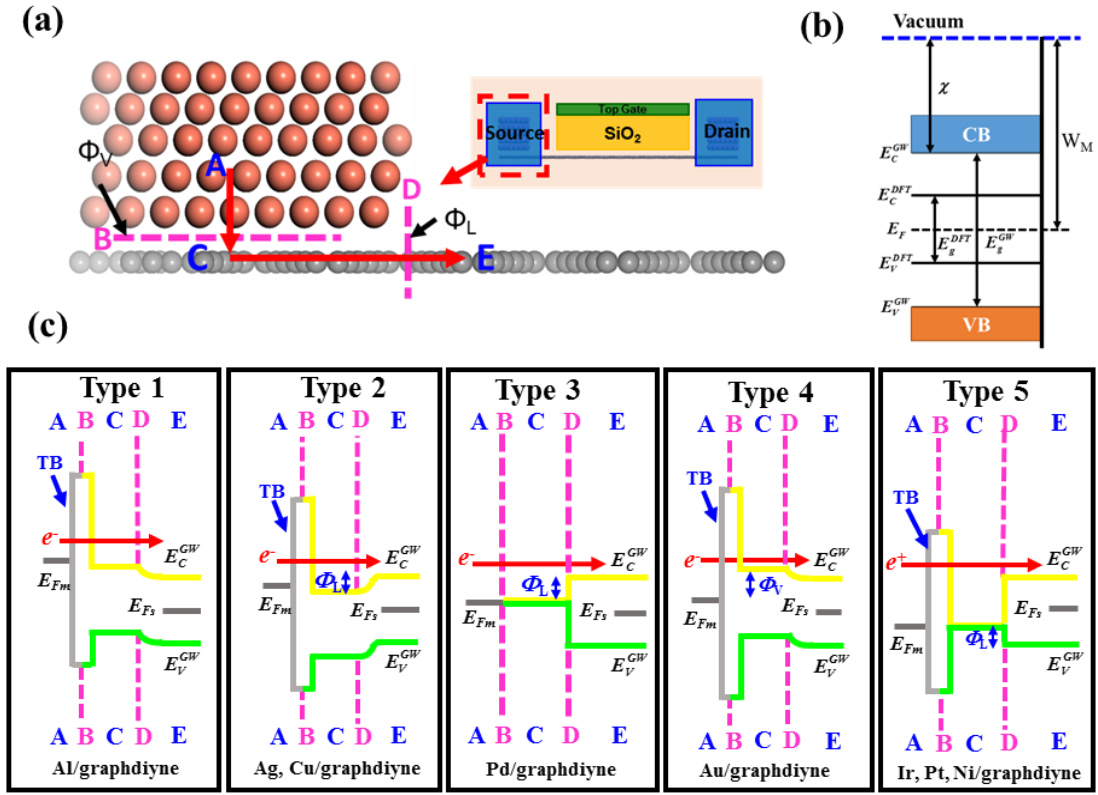

Figure 4 | (a) Schematic cross-sectional view of a typical metal contact to intrinsic graphdiyne. A, C, and E denotes three regions, while B and D are the two interfaces separating them. Red rows show the pathway (A→B→C→D→E) of electron or hole injection from contact metal (A) to the graphdiyne channel (E). Inset figure shows the source/drain contacts and the channel region in a typical top-gated FET. (b) Schematic illustration of the absolute band position with respect to the vacuum level by *GW* correction. (c) Five possible band diagrams of (a), depending on the type of metal. Examples are provided at the bottom of each diagram. $E_{Fm}$ and $E_{Fs}$ denote the Fermi level of absorbed system and channel graphdiyne, respectively.

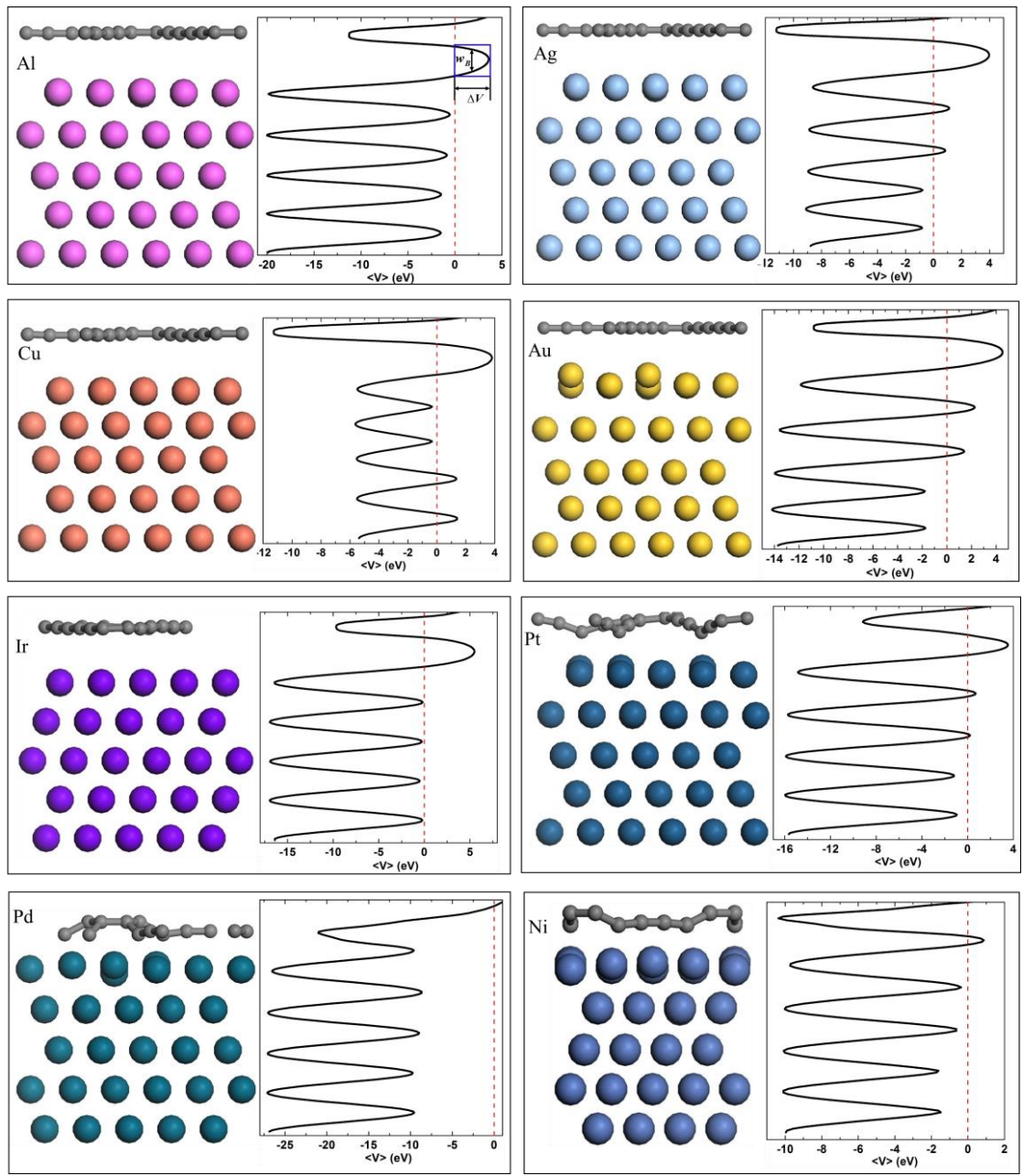

Figure 5 | Side view of optimized configuration and average electrostatic potential in planes normal to the interface of graphdiyne/Al, Ag, Au, Cu, Ir, Pt, Pd, and Ni systems, respectively. The Fermi level is set at V = 0 eV.

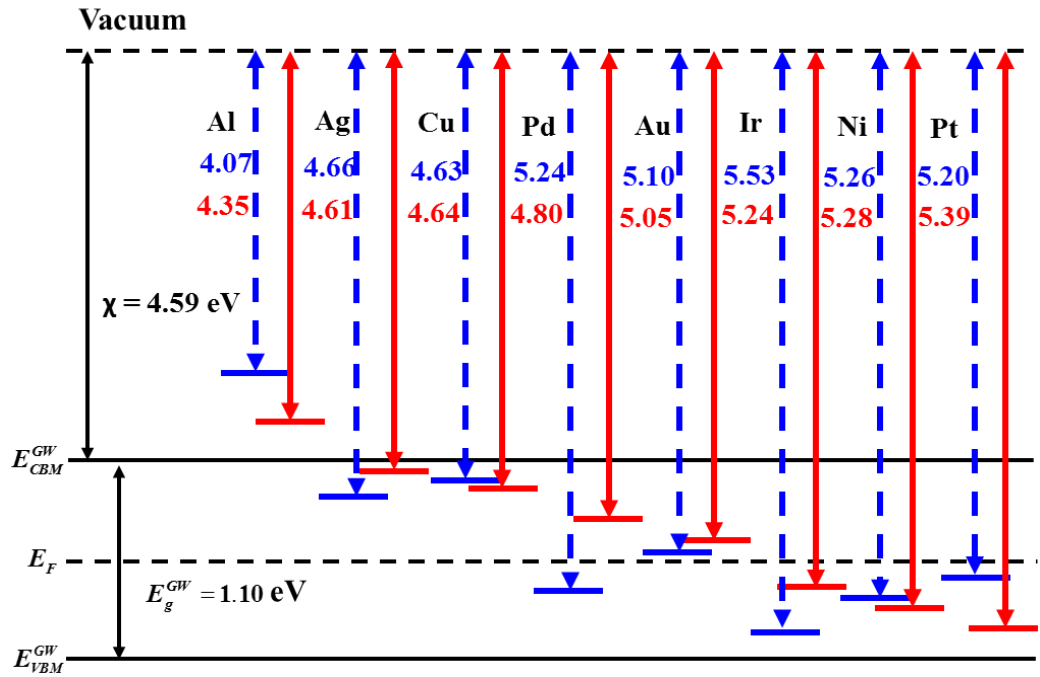

Figure 6 | Line-up of source work function with the *GW*-corrected electronic bands of channel graphdiyne. The blue dash line is the work function of pure metal, and the red solid line is the work function of contacted systems.

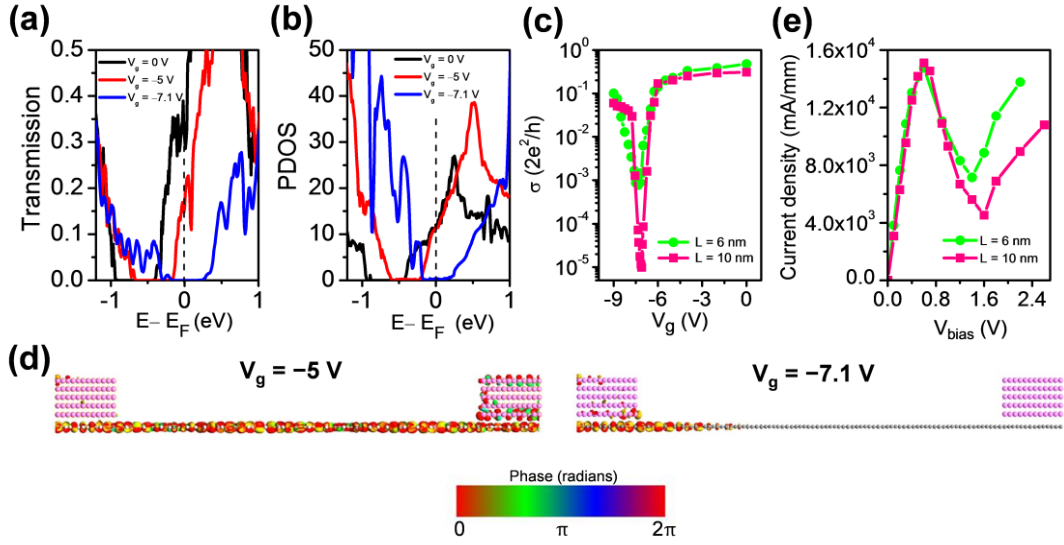

Figure 7 | Single-gated graphdiyne FET with Al electrodes: (a) Transmission spectra and (b) projected density of states of the channel under $V_g = 0$, −5, and −7.1 V in the graphdiyne FET with a channel length $L = 10$ nm. (c) Zero-bias transfer characteristics for $L = 6$ and 10 nm. (d) Transmission eigenstates at $E = E_F$ and at $k = (0, 0)$ for the on- ($V_g = -5$ V, left panel) and off-state ($V_g = -7$ V, right panel) with $L = 10$ nm. The isovalue is 0.2 a.u.. (e) Output characteristics of the 6 and 10 nm-channel-length graphdiyne FET under $V_g = 0$.

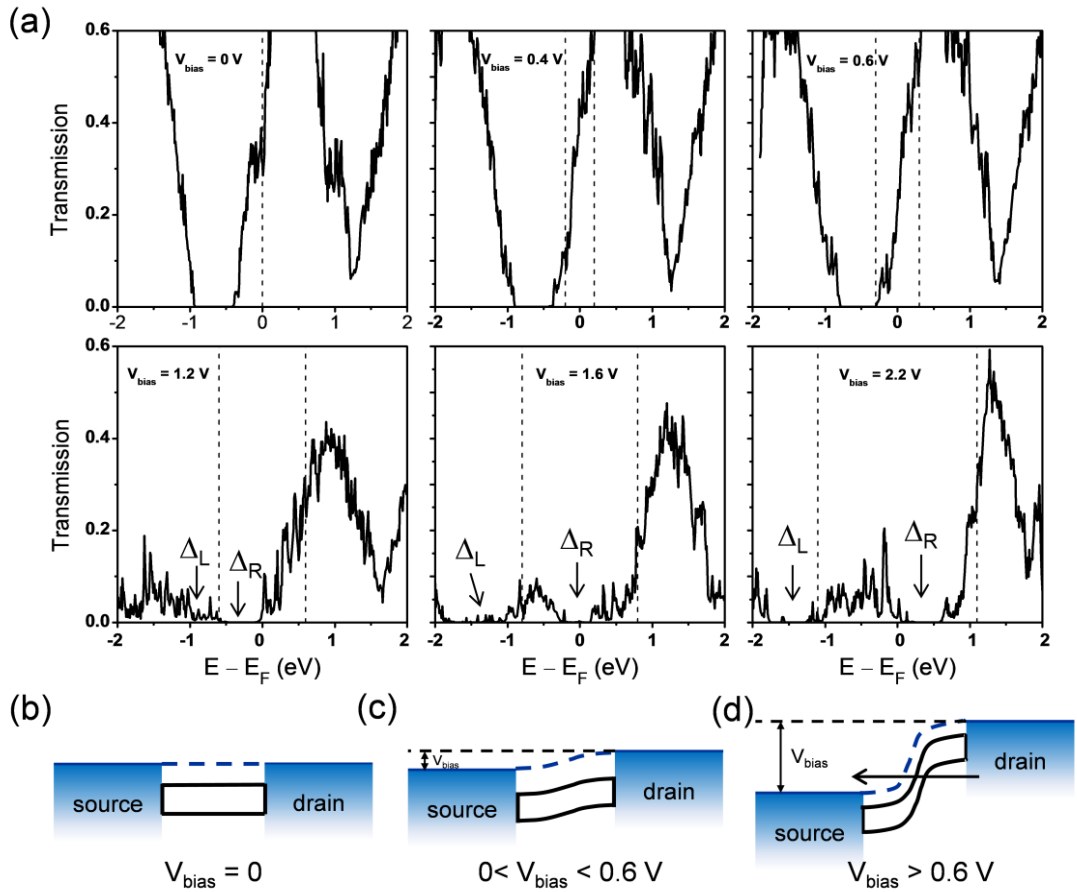

Figure 8 | (a) Transmission spectra and (b-d) band diagrams of the 10 nm-length-channel graphdiyne FET with Al electrodes under different $V_{bias}$ (negative drain bias) at $V_g = 0$. The black dashed vertical line indicates the bias window. $\Delta_{L(R)}$ denotes the transport gap induced by the band gap in the left (right) part of the channel near the source (drain) region.

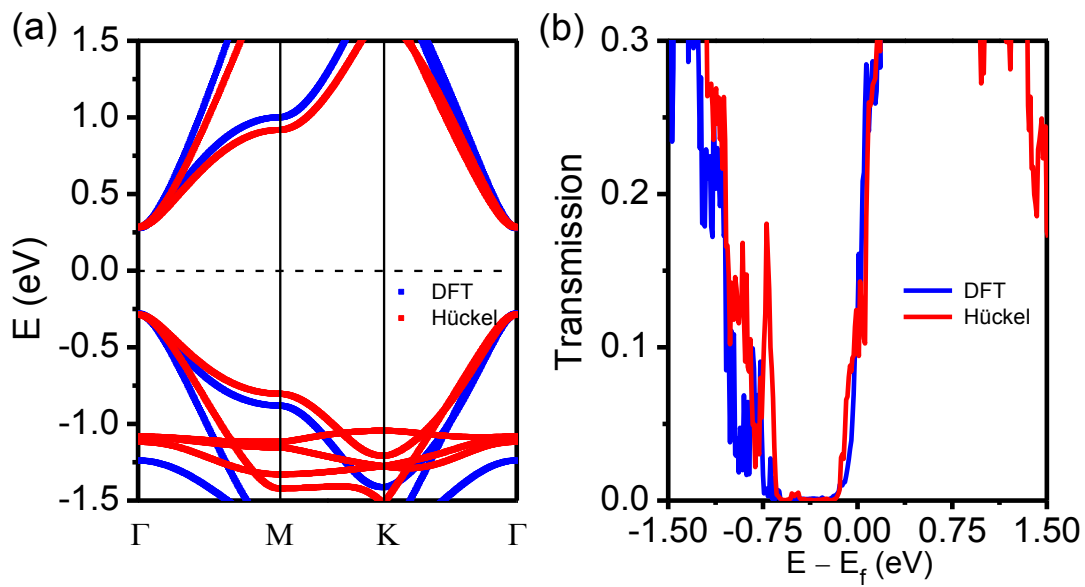

**Figure S1**. Comparison of (a) the band structure of graphdiyne and (b) the transmission spectra of the 6 nm-channel-length graphdiyne FET ($V_g = 0$ and $V_{bias} = 0$) calculated by DFT (blue) and SE (red) methods.